# Microscale magnetic resonance detectors: a technology roadmap for *in vivo* metabolomics[1]

Jan G. Korvink, Neil MacKinnon

Institute of Microstructure Technology, Karlsruhe Institute of Technology,
Hermann-von-Helmholtz-Platz 1, 76344 Eggenstein-Leopoldshafen, Germany

***Keywords***:  nuclear magnetic resonance, microcoils, hyperpolarisation, metabolomics, *Caenorhabditis elegans*

One of the great challenges in biology is to observe, at sufficient detail, the real-time workings of the cell. Many methods exist to do cell measurements invasively. For example, mass spectrometry has tremendous mass sensitivity but destroys the cell. Molecular tagging can reveal exquisite detail using STED microscopy, but is currently neither relevant for a large number of different molecules, nor is it applicable to very small molecules. For marker free non-invasive measurements, only magnetic resonance has sufficient molecular specificity, but the technique suffers from low sensitivity and resolution. In this presentation we will consider the roadmap for achieving *in vivo* metabolomic measurements with more sensitivity and resolution. The roadmap will point towards the technological advances that are necessary for magnetic resonance microscopy to answer questions relevant to cell biology.

1. **Introduction**

The basic functional subunit of living organisms is the cell. From an engineering point of view, the cell is a cubic volume of about $10\,\mu$m on a side (1 pl) that contains about $10^{10}$ active bio-molecules and water. Inside the cell, a number of functional compartments are found, such as a nucleus, or the mitochondria. Outside the cell, cells are the building blocks of ever more complex units, including vasculature, glands, organs, and structural elements. It is fairly straightforward to see that it is hard to draw conclusions about the functions of the larger units, and their interaction, without understanding the functioning of the cell and its constituents. So there is a strong motivation to reveal the dynamical behaviour of the smallest compartments of life, preferably in real time, and non-invasively, so that the observed behaviour closely resembles the behaviour within our bodies.

At the present time, it is not possible to make such measurements with any form of generality. (Because we are interested in non-invasive in vivo measurements, we only consider nuclear magnetic resonance - NMR - in this article.) As a result, it is currently necessary to enlarge the compartment size to a level where current instruments measure enough signal in a short enough time, and to declare this level as the practicable voxel and associated timestep. It is interesting to consider the various constraints on signal detection:

- *Metabolism*. The open access KEGG pathway database[1] describes the known metabolomic pathway maps for a range of organisms, linking and organising contributions from the scientific literature. Although largely converged, it is known that the metabolomic pathway map is not yet complete. Due to the large number of metabolites (and other molecular constituents), it is clear that the signal level due to a particular molecule in a cell will be very low, and only water and lipids will be present in sufficient quantity to enable detection at the single cell level. This means that, in order to observe a particular metabolomic pathway, or the flux through a particular

---

[1] This document is weakly abstracted from: *NMR microscopy for in vivo metabolomics, digitally twinned by computational systems biology, needs a sensitivity boost*, Jan G. Korvink, Vlad Badilita, Lorenzo Bordonali, Mazin Jouda, Dario Mager, Neil MacKinnon, to appear in Sensors and Materials (2018)

pathway, it is necessary to study a large collection of cells as an ensemble. The cells must be conditioned to respond to a stimulus in parallel, so as to make the response as synchronised as possible.

- *Organelle*. By definition, organelles are even smaller than cells, but may be slightly more homogeneous in molecular composition. Nevertheless, studying organelles may require even more averaging to produce detectable signals.
- *Cell variability*. An organism is made up of a very large number of different specialised cells. The boundaries of cell type are sometimes clear, and often not at all. For example in the liver, different cells are geometrically organised side-by-side at the single cell level. Since cells may vary dramatically in their metabolomic profile, enlarging the compartment must be performed with care if conclusions are still to be relevant.
- *Time*. NMR experiments take a finite amount of time. Signal enhancement is often performed by averaging over multiple detection cycles, which means that the signal is averaged over a time span. This time span is the limit of dynamics that the NMR experiment can reveal, regardless of the dynamics of the biological system.
- *Space*. It is possible to encode the spatial origin of a particular NMR signal (this is what magnetic resonance imaging is all about). The higher the spatial resolution, the smaller the number of molecules that participate in the experiment. Thus spatial resolution is gained at the expense of temporal and spectral resolution.
- *Spectrum*. The NMR experiment stores a certain amount of electromagnetic energy into the nuclear spin system of the sample's molecules through the absorption of radiofrequency excitation. During NMR detection, the molecules relax via numerous spin pathways to thermodynamic equilibrium. The emitted energy is collected by NMR detectors in frequency bands that correspond to various nuclei. The amount of energy that corresponds to a particular resonance peak is proportional to the area under that peak. Any randomness in the signal (noise) will limit the information that can be derived from the peak and hence will introduce some uncertainty. The more distinguishable constituents that are excited in a particular experiment, the more peaks there will be.
- *Polarisability*. Due to the low energy of radiofrequency radiation, NMR is inherently an insensitive technique. The Boltzmann distribution $\exp(-E/kT)$ defines the polarisation degree of a nuclear spin, and the low energy level of $6.626 \times 10^{-34}$ J/Hz reveals that NMR signals are close to thermally induced noise, and justifies the use of very strong magnets with which to perform spectroscopy.
- *Proximity*. NMR is usually detected by induction. It is also possible to detect NMR signals using dipolar coupling to either a nanoscale magnet, or to a nanoscale nitrogen vacancy. All three detection techniques benefit enormously by moving the detector closer to the spin. For non-invasive detection, only induction is suited since it also works at larger distances up to 1 m.

Because of the many constraints, also with their interdependencies, it is necessary to select a specific problem to work on. At the MEMS scale, it is pragmatic to focus one's attention on an organism that has a MEMS-like length scale, in which the nematode *Caenorhabditis elegans* fits very well.[2] The many other reasons to do so lean on those which Sydney Brenner originally used to justify selecting the worm as a basis for rational genomics research:[3]

- *Standardisation*. The embryonic development of *C. elegans* follows an identical map, so that cellular predecessors are exactly known. Responses are also largely programmable, so that a worm colony can be reasonably synchronised.

- *Tractability*. With only ~$10^3$ cells, and 302 neurons, localisation is readily related to function, and to behaviour.
- *Transparency*. Phenotypes are not always optically distinguishable, so that molecular phenotyping takes on a particular significance.
- *Physical dimensions and other practicalities*. The worm has about the dimensions of practical microfluidics, and represents approximately the smallest organism size where inductive NMR still has a scaling advantage. It has a very short lifecycle, allowing fast experimentation, and produces large amounts of progeny. The worms can be maintained at -80 °C, and revived easily.

Clearly, in order to tackle the metabolomics of such a small organism, the NMR experiment must be driven towards more sensitivity. Next, we consider some of the measures that can be taken.

2. **A roadmap of measures**

NMR microscopy clearly requires an overall sensitivity boost. Through an analysis of the constraints presented in the previous section, we can now consider technical measures that have been identified as promising:

- *Signal localisation*. A few conventional strategies are available for the strong localisation of NMR signals. The NMR signal, essentially present across the entire region of interest, can be spatially encoded by manipulating its phase through the application of magnetic field gradients. Because the NMR signal is of a quadrature nature (vectorial, with amplitude and phase), only two dimensional signals can be encoded in this fashion. The third dimension is usually *a priori* selected using a gradient field as well. By applying very strong gradients, the localisation can be reach a few micrometres, but further miniaturisation is currently not possible. Through the selective excitation of spins, further localisation can be achieved. Because the wavelength of NMR signals is very large (typically 10 cm - 1 m), RF localisation is not possible at microscopic levels.
- *Sample localisation*. *In vivo* implies that the organism of interest is alive and well, which means that, in general, it will also be motile. For metabolomics, it is important that the organism "feels" free to a certain degree, or else the metabolomic response will be that of a pure stress reaction. A potential conflict of requirements thus arises, because accurate measurements at the limit-of-detection require a stationary sample, and natural response implies a lack of constraints. Technology must therefore be developed that verifiably reduces organism stress yet restrains movement during detection.
- *Signal enhancement*. Despite all technical efforts, NMR at high field using the smallest available detectors will suffice for only the most abundant of metabolite. For metabolomics, more than one metabolite must be detected, so that the first conclusions about a pathway can be drawn from measurement. Concentration limits-of-detection is the appropriate quality measure, and despite small sample size, low metabolite concentrations of at least $1\,\mu$M must be achieved for meaningful biological experimentation.Thus non-equilibrium polarisations must be aimed for, and a number of potential candidates are currently on the horizon. Ground state para-hydrogen is a convenient source of polarisation which can be transferred from the hydrogen molecule to a substrate either via a catalytic process called SABRE,[4] or via a hydrogenation reaction called PHiP.[5] For the life sciences, relevant experiments could be to pass hyperpolarized molecules into the organism along the food chain, because the know catalysts for SABRE are toxic, and cannot be performed *in vivo*. Another hyperpolarization pathway is the use of electron polarisation of a radical with unpaired electron, generated for example with the help of microwaves, or by light in

- a diamond film equipped with NV centres.[6] The electron spin polarisation can be transferred to a sample's protons by Overhauser or solid effect, assisted by spin diffusion. Also here, the food chain is probably the most likely pathway into a small organism.
- *Detection*. NMR signal detection must take place at the highest possible signal-to-noise ratio. An immediate consequence for inductive measurements is the need for very high microcoil filling factors, and very high coil efficiencies measured in $B_1/\hat{I}$. Thus solenoidal detectors are needed that have inductors of the order of 100 $\mu$m diameter, because saddle coils and Helmholtz coils are less efficient.[7-14]
- *Spectroscopy*. *In vivo* spectra are very crowded compared to typical laboratory scale experiments, because purification of the sample is not possible, thus lipids and culture medium of the organism all contribute to clouding the measurement. As a result, exceptionally good chemometric methods are necessary. Selective excitation, coupled with solvent suppression methods, are necessary but not sufficient measures. Orthogonal spectral spaces must be exploited to unravel the mixtures and determine quantitative (in a relevant sense) signals.[15] Metabolites must be measured in their pure form, to determine possible similarities, and procedures must be characterised on standard mixtures to determine limits of separation, and the interplay between limit-of-detection and separability.

3. **Conclusions**

*In vivo* metabolomics is a **grand challenge** for biology. Once it is achieved, tremendous experimental opportunities will open up for experimentation, and for the clarification of disease progression, because metabolomic rates reveal the dynamics of all intra-cellular processes. Currently, microscale NMR metabolomics is the only method that has the potential to be both hands-free and *in vivo*. Due to the lack of sensitivity,[16] it is also a grand challenge for NMR science and technology. In the presentation accompanying this abstract, I will explore the technological methods that we have developed on this roadmap, and point the way ahead.

**Acknowledgements**

The work leading to this paper was primarily supported by the European Research Council (ERC) under grant number 290586 (NMCEL). Additional support was provided by the Deutsche Forschungsgemeinschaft (DFG), in the framework of the German Excellence Initiative under grant number EXC 1086 (BrainLinks-BrainTools), and grant number KO1883/23-1 (RUMS). The first author is greatly indebted to his hard-working team and collaborators who supported or contributed the numerous results that are summarized in this publication. Besides the co-authors, these include: Shyam Adhikari, Natalia Bakhtina, Erwin Fuhrer, Andreas Greiner, Oliver Gruschke, Jürgen Hennig, Jens Höfflin, Jan Hövener, Robert Kamberger, Ronald Kampmann, David Kauzlaric, Sebastian Kiss, Mona Klein, Kai Kratt, Robert Meier, Markus Meissner, Ali Moazenzadeh, Nikolaus Nestle, Kirill Poletkin, Herbert Ryan, Pedro Silva, Christoph Trautwein, Marcel Utz, Ulrike Wallrabe, Nan Wang, Peter While, Maxim Zaitzev.